\documentclass[aps, prl, twocolumn, superscriptaddress, longbibliography]{revtex4-2}

\usepackage{graphicx}
\usepackage{dcolumn}
\usepackage{bm}
\usepackage{hyperref}
\usepackage{todonotes}
\usepackage{color}
\usepackage{glossaries}
\usepackage{sidecap} 
\usepackage{orcidlink}
\usepackage[no-test-for-array]{nicematrix}

\newacronym{RIXS}{RIXS}{Resonant inelastic x-ray scattering}
\newacronym{XAS}{XAS}{x-ray absorption spectroscopy}
\newacronym{ARPES}{ARPES}{angle-resolved photoemission}
\newacronym{ED}{ED}{exact diagonalization}
\newacronym{AIM}{AIM}{Anderson impurity model}
\newacronym{vdW}{vdW}{van der Waals}
\newacronym{DFT}{DFT}{density functional theory}
\newacronym{ESRF}{ESRF}{European Synchrotron Radiation Facility}
\newacronym{TFY}{TFY}{total fluorescence yield}

\newacronym{1D}{1D}{one-dimensional}
\newacronym{2D}{2D}{two-dimensional}

\newacronym{RLU}{RLU}{reciprocal lattice units}

\begin{document}

\title{Observation of anisotropic dispersive dark exciton dynamics in CrSBr}

\author{J. Sears\,\orcidlink{0000-0001-6524-8953}}
\affiliation{Department of Condensed Matter Physics and Materials Science, Brookhaven National Laboratory, Upton, New York 11973, USA}
\author{B. Zager\,\orcidlink{0000-0001-7436-4040}}
\affiliation{Department of Condensed Matter Physics and Materials Science, Brookhaven National Laboratory, Upton, New York 11973, USA}
\affiliation{National Synchrotron Light Source II, Brookhaven National Laboratory, Upton, New York 11973, USA}
\author{W. He\,\orcidlink{0000-0003-3522-3899}}
\altaffiliation[Current address: ]{Stanford Institute for Materials and Energy Sciences, SLAC National Accelerator Laboratory, Menlo Park, CA 94025}
\affiliation{Department of Condensed Matter Physics and Materials Science, Brookhaven National Laboratory, Upton, New York 11973, USA}
\author{C. A. Occhialini\,\orcidlink{0000-0002-2133-5155}}
\affiliation{National Synchrotron Light Source II, Brookhaven National Laboratory, Upton, New York 11973, USA}
\affiliation{Department of Physics, Columbia University, New York, NY, USA}
\author{Y. Shen\,\orcidlink{0000-0003-4697-4719}}
\altaffiliation[Current address: ]{Beijing National Laboratory for Condensed Matter Physics, Institute of Physics, Chinese Academy of Sciences, Beijing 100190, China; School of Physical Sciences, University of Chinese Academy of Sciences, Beijing 100049,
China}
\author{M. Lajer,\orcidlink{0000-0002-1168-8598}}
\affiliation{Department of Condensed Matter Physics and Materials Science, Brookhaven National Laboratory, Upton, New York 11973, USA}

\author{J.~W.~Villanova\,\orcidlink{0000-0003-1595-3851}}
\altaffiliation[Current address: ]{Department of Physics and Astronomy, Middle Tennessee State University, Murfreesboro, Tennessee 37132, USA}
\author{T. Berlijn\,\orcidlink{0000-0002-1001-2238}}
\affiliation{Center for Nanophase Materials Sciences, Oak Ridge National Laboratory, Oak Ridge, Tennessee 37831, USA}

\author{F. Yakhou-Harris\,\orcidlink{0000-0002-6798-2286}}
\author{N. B. Brookes\,\orcidlink{0000-0002-1342-9530}}
\affiliation{European Synchrotron Radiation Facility (ESRF), B.P. 220, F-38043 Grenoble Cedex, France}

\author{D.~G.~Chica\,\orcidlink{0000-0001-8616-9365}}
\author{X. Roy\,\orcidlink{0000-0002-8850-0725}}
\affiliation{Chemistry Department, Columbia University, New York, NY 10027, USA}

\author{E.~Baldini\,\orcidlink{0000-0002-8131-9974}}
\affiliation{Department of Physics, The University of Texas at Austin, Austin, Texas, USA, 78712}

\author{J. Pelliciari\,\orcidlink{0000-0003-1508-7746}}
\author{V. Bisogni\,\orcidlink{0000-0002-7399-9930}}
\affiliation{National Synchrotron Light Source II, Brookhaven National Laboratory, Upton, New York 11973, USA}

\author{S.~Johnston\,\orcidlink{0000-0002-2343-0113}}
\affiliation{Department of Physics and Astronomy, The University of Tennessee, Knoxville, Tennessee 37966, USA\looseness=-1}
\affiliation{Institute of Advanced Materials and Manufacturing, The University of Tennessee, Knoxville, Tennessee 37996, USA\looseness=-1}

\author{M. Mitrano\,\orcidlink{0000-0002-0102-0391}}
\affiliation{Department of Physics, Harvard University, Cambridge, Massachusetts 02138, USA\looseness=-1}

\author{M. P. M. Dean\,\orcidlink{/0000-0001-5139-3543}}
\email{mdean@bnl.gov}
\affiliation{Department of Condensed Matter Physics and Materials Science, Brookhaven National Laboratory, Upton, New York 11973, USA}

\date{\today}

\begin{abstract}
Many-body excitons in CrSBr are attracting intense interest in view of their highly anisotropic magneto-optical coupling and their potential for novel optical interfaces within spintronic and magnonic devices. Characterizing the orbital character and propagation of these electronic excitations is crucial for understanding and controlling their behavior; however, this information is challenging to access. Ultra-high resolution resonant inelastic x-ray scattering is a momentum-resolved technique that can address these crucial questions. We present measurements collected at the Cr $L_3$-edge which show a rich spectrum of excitations with a variety of spin-orbital characters. While most of these excitations appear to be localized, the dispersion of the lowest energy dark exciton indicates that it is able to propagate along both the $a$ and $b$ directions within the planes of the crystal. This two-dimensional character is surprising as it contrasts with electrical conductivity and the behavior of the bright exciton, both of which are strongly one-dimensional. The discovery of this propagating dark exciton highlights an unusual coexistence of one- and two-dimensional electronic behaviors in CrSBr.
\end{abstract}

\maketitle

\textit{Introduction.}---The weak dielectric screening intrinsic to \gls*{2D} \gls*{vdW} materials promotes the formation of strongly bound excitons, quasiparticles consisting of a bound electron and hole pair. Excitons often dominate the optical response in these materials and have for this reason been intensively studied for potential optoelectronics applications \cite{Wang2018colloquium, Mueller2018, Wilson2021review}. Optically forbidden dark excitons are highly sensitive to the electronic states of their host material \cite{Mueller2018, Molas_2017}, and are of interest for applications exploiting exciton coherence  \cite{Zhu2024}. More recently, a variety of magnetic \gls*{vdW} materials have been reported \cite{Burch2018magnetism, Wang2022magnetic} with the potential to support excitons with intertwined electronic and magnetic properties. \gls*{RIXS} is a versatile measurement technique that can detect these various types of electronic excitation, and provide information on their spin and orbital character not available from other techniques. \gls*{RIXS} also provides information about the exciton dispersion at nonzero momentum transfer, therefore probing the propagation of these excitations in the material \cite{Mitrano2024exploring}. This type of in-depth characterization will be valuable for understanding and tuning functional excitons for incorporation into devices.

CrSBr is a \gls*{vdW} material that was recently found to be magnetic down to the monolayer limit \cite{Lee2021magnetic}, and also possesses several other desirable properties for embedding into electronic and magnetic devices. It is semiconducting with a direct band gap of $\sim1.8$~eV, air-stable, and has a relatively high ordering temperature of $T_N\approx 132$~K \cite{Goser1990magnetic, Telford2020magnetoresistance, Scheie2022spin, Ziebel2024CrSBr, Smolenski2025large}. Optical studies have identified a bright exciton at 1.37~eV, which shows a strong coupling with magnetism \cite{Wilson2021interlayer, Bae2022exciton}. This discovery has prompted a great deal of research investigating the potential for optical control and detection of electronic and magnetic quasiparticle excitations in CrSBr \cite{Bae2022exciton, Diederich2023tunable, Dirnberger2023magneto}.

\begin{figure*}
\includegraphics[width=\textwidth]{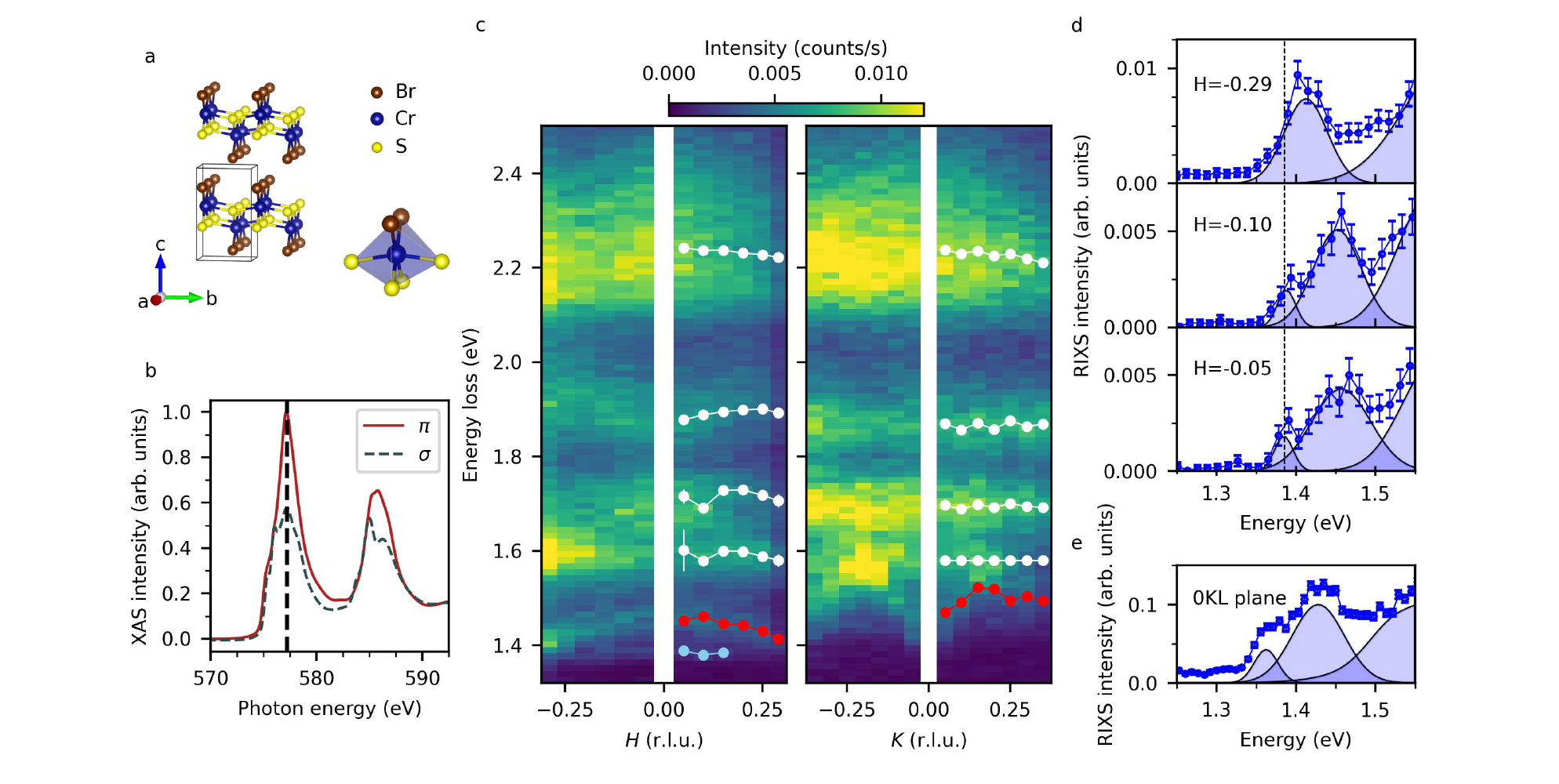}
\caption{Anisotropic dispersive exciton behavior in CrSBr (a) Crystal structure. The gray cuboid shows the unit cell and the translucent blue octahedron shows the Cr coordination. (b) X-ray absorption measured in \gls*{TFY} mode as a function of incident photon energy, showing peaks at the Cr $L_3$ and $L_2$ edges. The data was collected in the $H0L$ plane, with an incident angle of $\theta_i = 45^{\circ}$ and a scattering angle of $2\Theta = 90^{\circ}$. The dashed vertical line indicates the energy at which the \gls*{RIXS} spectra in (c) were collected. $\sigma$ polarization indicates vertical linear polarization of the incoming beam along the sample $b$-axis, and $\pi$ horizontal linear polarization in the sample $a-c$ plane.  (c) Exciton propagation along the $H$ and $K$ directions of the Brillouin zone. The peak positions extracted from the fit are plotted on one half of the data in each plane, showing non-dispersive behavior in most of the excitons (white), as well as dispersive propagating behavior in one of the low-energy excitons (red). A non-dispersive low energy feature, which we identify as the previously reported optically active exciton, is visible at low momentum transfers in the $H0L$ plane (blue). The $H$ dependent measurements were collected at $K=0$, with the beam polarization in the $H0L$ scattering plane. The $K$ dependent measurements were collected at $H=0$, with the photon polarization along the $H$ axis perpendicular to the $0KL$ scattering plane. (d) Line cuts showing raw spectra and Gaussian fits to the low energy excitations in the $H0L$ plane, demonstrating that two peaks can be resolved in the 1.3 to 1.5~eV energy range. The vertical dashed line marks the position of the 1.38(1)~eV feature. (e) Additional data measured with the sample in the $0KL$ plane at $K=0$, demonstrating that the weak feature can also be resolved at sufficiently low momentum transfers. 
}
\label{fig1}
\end{figure*}

The bright exciton in CrSBr is understood to have a delocalized Wannier character and to arise from hybridization between the Cr $3d$ and S and Br $p$ states \cite{Wilson2021interlayer}. Although prior studies have reported a number of optically bright features in the 1.3-1.4 eV energy range, these result from coupling either to phonons or polariton modes \cite{Lin2024phonon, Wang2023magnetically, Shao2025magnetically}. The bare exciton energy has been found to be 1.37~eV in \cite{Wang2023magnetically}. Photoluminesence measurements of the exciton show a striking anisotropy, with the exciton only visible when the light is polarized along the crystallographic $b$-axis \cite{Wilson2021interlayer, Wu2022quasi} and has been termed a \gls*{1D} exciton. This is a consequence of the highly anisotropic band structure, which results in quasi-\gls*{1D} electronic conduction along the $b$ axis and localization along the $a$ axis. The \gls*{1D} electronic behavior is counter-intuitive, but understandable through a careful inspection of the crystal structure \cite{Goser1990magnetic} shown in Fig.~\ref{fig1}(a), which is made up of apparently well-connected bilayers of distorted CrS$_4$Br$_2$ octahedra. The electronic anisotropy is due to the Cr-ligand-Cr bond angles, which are $\sim$90$^{\circ}$ along $a$ and $\sim$180$^{\circ}$ along $b$. The quasi-\gls*{1D} electronic conduction occurs along the nearly straight chains of Cr and S running along the $b$ direction \cite{Wu2022quasi, Klein2023bulk}.

Here we present the first use of \gls*{RIXS} to determine the electronic character and propagation of excitons in CrSBr. We observed an exciton at 1.38(1)~eV consistent with the previously reported bright exciton at 1.37~eV \cite{Wang2023magnetically} and found that how this feature changes with momentum is consistent with that expected for a \gls*{1D} Wannier exciton following prior description of this mode as a \gls*{1D} quasiparticle \cite{Klein2023bulk}. We also detected several previously unreported excitations with higher energies up to 2.5~eV, including a dark exciton at 1.45 eV. While the higher-energy excitations are non-dispersing, indicating localized electronic excitations, the lowest-energy dark exciton at 1.45 eV shows more exotic behavior. This excitation shows a strong temperature dependence and an unusual \gls*{2D} dispersion more consistent with local exchange hopping than with a Wannier exciton. The \gls*{2D} nature of this propagating dark exciton contrasts with the generally \gls*{1D} electronic behavior, highlighting the complex interplay between the \gls*{1D} and \gls*{2D} phenomena in CrSBr.

\begin{figure}
\includegraphics[width=\columnwidth]{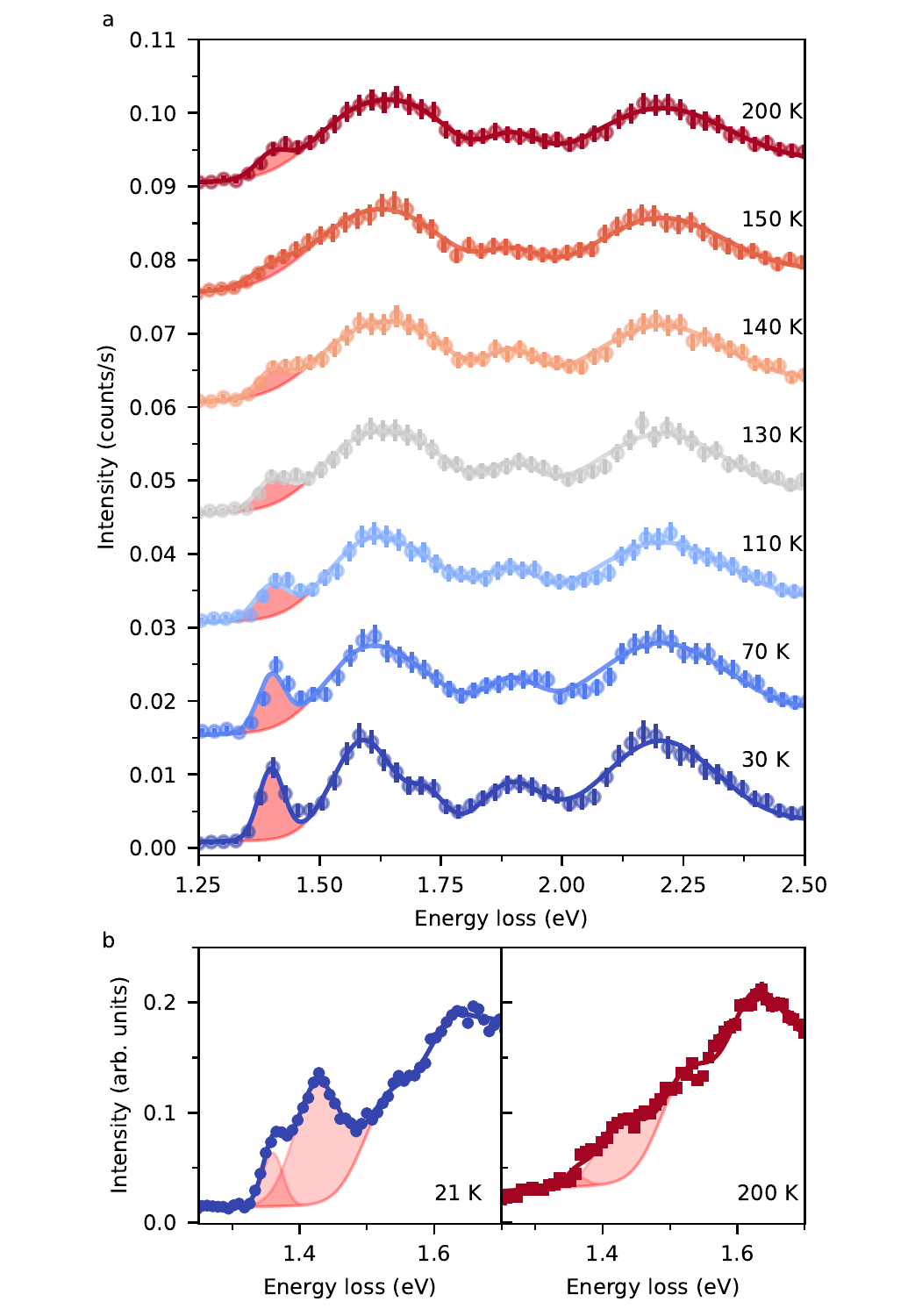}
\caption{Low-temperature emergence of excitons in CrSBr (a) \gls*{RIXS} spectra with varying temperature, collected at 577.2~eV ($L_3$ edge) at $\theta_i=9.4^{\circ}$ and $2\Theta=149^{\circ}$. The prominent excitation is visible at $\sim$1.4~eV is the dark dispersive exciton and is highlighted by the pink shading at low temperature. This feature becomes weaker and broader with increasing temperature. The spectra were fit with five Gaussian functions with temperature-dependent widths and intensity, and constant energies. (b) Temperature dependence measured at lower momentum transfer to resolve the two low energy peaks. The optically active and the dark dispersive excitons both show a similar decrease in intensity at higher temperatures above $T_N$. This data was collected at $\theta_i=68.5^{\circ}$ and $2\Theta=150^{\circ}$. Data in both panels was collected in the $H0L$ plane with a $\pi$-polarized beam.}
\label{fig2}
\end{figure}

\textit{Methods.}---Single crystals of CrSBr were synthesized and aligned as described in the Supplemental Material Sec.~I \cite{supp}.  Cr $L_3$-edge \gls*{RIXS} measurements were performed at the ID32 beam line of the \gls*{ESRF} \cite{Brookes2018beamline}. $\sigma$ polarization indicates vertical linear polarization of the incoming beam (perpendicular to the scattering plane), and $\pi$ denotes horizontal linear polarization (in the scattering plane). Unless otherwise specified, a sample temperature of 30~K was used. The spectrometer was operated with a resolution of 30~meV for the reciprocal space maps and temperature dependence. Additional spectra (Fig.~\ref{fig1}(e) and Fig.~\ref{fig2}(b)) were collected at the SIX 2-ID beamline of the National Synchrotron Light Source II. The spectrometer was operated with a high energy resolution of $28$~meV. The CrSBr data were interpreted using the \textsc{EDRIXS} software as described in the End Matter \cite{Mitrano2024exploring, Wang2019EDRIXS, EDRIXS}.

\textit{Results and interpretation.}---We used \gls*{RIXS} to measure the electronic excitations and their momentum-dynamics in CrSBr, with the incident photon energy tuned to the peak of the \gls*{XAS} spectrum at the chromium $L_3$ edge. The Cr $L_{2,3}$ \gls*{XAS} measurements collected with horizontal ($\pi$) and vertical ($\sigma$) incident beam polarizations are shown in Fig.~\ref{fig1}(b). These data were collected below the magnetic ordering temperature and so show dichroism due to crystallographic and magnetic symmetry breaking.

The \gls*{RIXS} spectra collected at the $L_3$ edge show several features between 1.3 and 2.5~eV energy loss, corresponding to different electronic excitations within the chromium $d$ orbitals and hybridized ligand (S and Br) states. In Fig.~\ref{fig1}(c) we show the momentum dependence of these excitations as a function of the $H$ and $K$ reciprocal space directions (CrSBr is a highly \gls*{2D} material, so no variation is expected in $L$ \cite{Scheie2022spin}). No features were observed in the \gls*{RIXS} spectra below 1.3~eV other than a shoulder on the elastic line, which is likely to arise from low energy ($<$100~meV) magnons and phonons \cite{Scheie2022spin}. At energies higher than 2.5~eV, no sharp features were seen.

Within the 1.3 to 2.5~eV range, we have fit the spectrum at each momentum transfer with a collection of Gaussian peaks as shown in Supplemental Material Sec.~II \cite{supp}. The peak positions derived from these fits are plotted on the right-hand side of each reciprocal space map. At most momentum transfers, we fit the spectra with five Gaussian peaks, however an additional weak peak was seen at low energy in the low momentum transfer region of the $H$ direction reciprocal space map. The position of this peak (1.38(1)~eV) is marked by the blue points in Fig.~\ref{fig1}(c) and agrees well with the 1.37~eV energy reported for the exciton in optical measurements \cite{Wang2023magnetically} and so we identify this as the optically active excitation which has been extensively studied in the literature.  Figure~\ref{fig1}(d) shows several spectra from this region as line cuts, which show more clearly the presence of this weak feature distinct from the higher energy excitations. The 1.38(1)~eV peak is not seen in the $K$ direction map, which is expected behavior for a band-edge excitation which would quickly disperse to higher energies in this direction due to the band anisotropy. It would then not be observable in our measurements since it would overlap with the higher energy excitations. We have also collected an additional spectrum at $K=0$, plotted in Fig.~\ref{fig1}(e), showing that this excitation can be resolved in the $0KL$ scattering plane geometry at sufficiently small momentum transfers.

\begin{figure}[!hbtp]
\includegraphics[width=1\columnwidth]{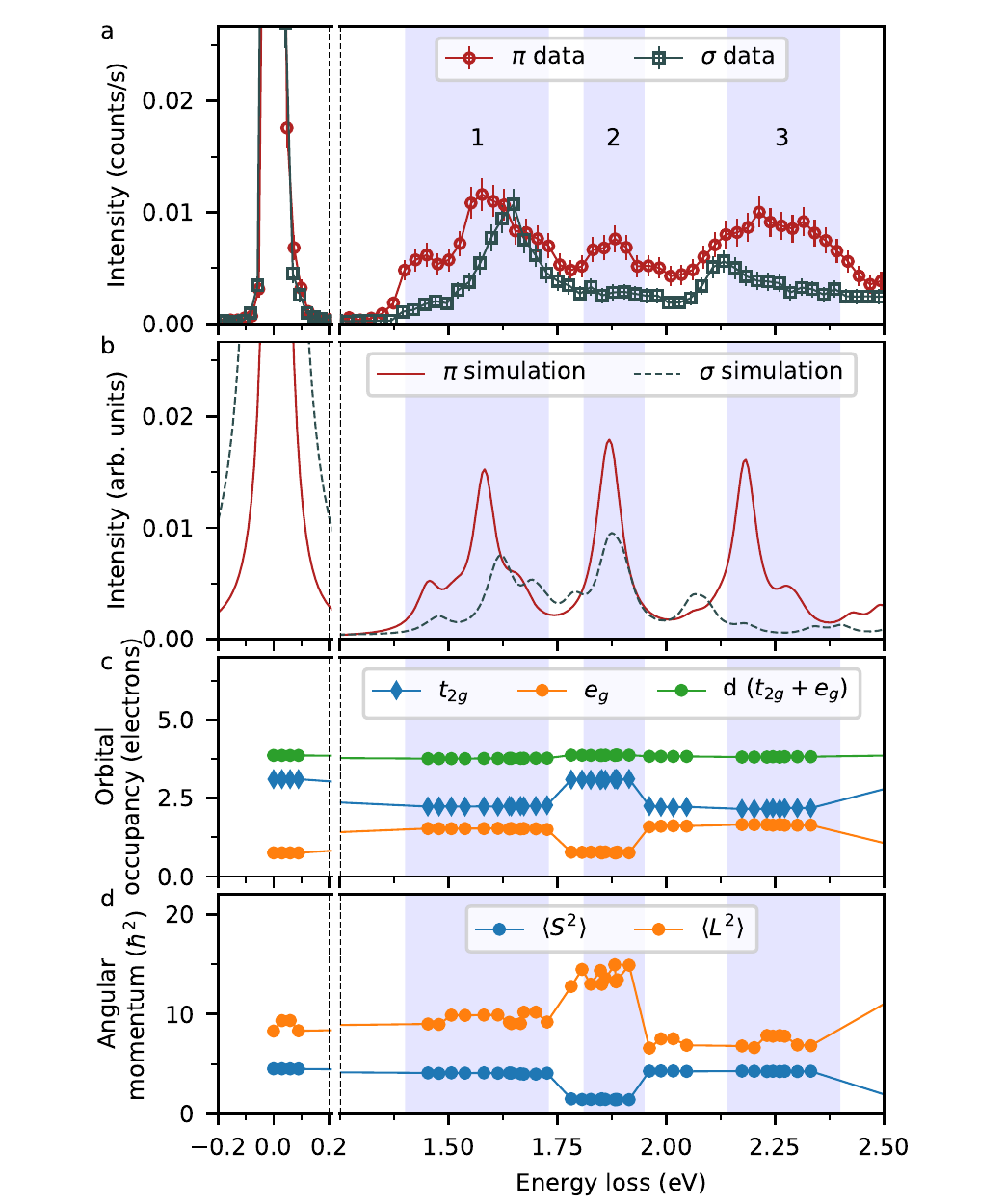}
\caption{Electronic character of the CrSBr excitons. (a) Low-temperature (30~K) \gls*{RIXS} spectra collected in the $H0L$ plane at $\theta_i=15^{\circ}$, $2\Theta=90^{\circ}$ and 577.4~eV. (b) \gls*{AIM} simulations that capture the main structures and polarization dependence observed in \gls*{RIXS}. (c),(d) Plot the orbital occupancies and spin and orbital angular momentum expectation values squared, denoted $\langle S^2\rangle$ and $\langle L^2\rangle$, respectively. We identify three manifolds of excitations: $t_{2g}\rightarrow e_g$ excitations below 1.75~eV, high spin-to-low-spin excitations from 1.75-2~eV, and a second manifold of $t_{2g}\rightarrow e_g$ excitations with reduced orbital angular momentum $L$ above 2~eV.}
\label{fig3}
\end{figure}

We find that most of the higher energy excitations do not show unambiguous dispersive behavior \footnote{We note that some of the higher energy features around 1.6~eV, such as those around 2.2~eV, show apparent changes in the peak center of mass with $H$ and $K$. In these cases, the magnitude of the changes is smaller than the peak width. Given that these peaks contain substructures which can vary in intensity at different values of $H$ and $K$, we do not consider our observations to be unambiguous evidence of true dispersion.}, with the exception of the low energy feature (1.46~eV at zero momentum transfer) marked by the red points. This excitation shows an intriguing asymmetry, dispersing downwards along $H$ and upwards along $K$ by a similar amount ($\sim 50$~meV). This type of dispersion is incompatible with a conventional particle-hole picture of the exciton, which would require a far larger exciton dispersive bandwidth (of order eV in magnitude) along $b$ due to the larger conduction band dispersion in this direction \footnote{Our \gls*{DFT} calculations in Fig.~S1 show a huge difference in the conduction band dispersion along $H$ and $K$. Along $\Gamma \rightarrow Y$ the conduction band disperses by 2~eV, whereas $\Gamma \rightarrow X$ is almost flat.}, but is compatible with an exciton that moves via exchange processes. This can be rationalized in view of bond-angle-dependent cancellation between different superexchange processes \cite{Kanamori1959superexchange}. This type of cancellation is thought to be the mechanism behind the relatively isotropic magnon bandwidth \cite{Scheie2022spin}. Since this dispersive excitation has not been reported in optical measurements, we will refer to this as a ‘dark’ exciton. We estimate the energy of the dark dispersive exciton to be 1.46(1)~eV at zero momentum transfer, somewhat larger than the 1.38(1)~eV energy observed for the bright exciton. The peak width was found to be 38(2)~meV, slightly increased from the 30~meV resolution likely due to lifetime broadening. The peak width of the 1.38(1)~eV bright exciton was resolution limited.

In Fig.~\ref{fig2}(a) we show the temperature dependence at the momentum transfer $\vec{q}=(-0.29, 0, 0.28)$ reciprocal lattice units from the base temperature of 30~K up to 200~K, well above the magnetic ordering temperature of 132~K. We find that the dark dispersive excitation shows the largest change, nearly disappearing at the highest temperatures \footnote{In the temperature dependence in Fig.~\ref{fig3}, the data at 150~K shows a slightly less intense exciton than is expected within the overall trend. We believe this does not constitute a physically meaningful change.}. While the 1.37 eV optically active exciton is not visible at this momentum transfer, we have collected additional spectra at smaller momentum transfer and found that the bright exciton shows a similar behavior at high temperature. These additional spectra are shown in Fig.~\ref{fig2}(b). While the higher energy excitations become generally broader and weaker, they do not undergo the same loss in spectral weight. 

To gain more insight into the nature of the electronic excitations, we simulated the \gls*{RIXS} spectra using a cluster model as described in the End Matter. \gls*{RIXS} has unique advantages for studying nominally optical dipole forbidden many body excitons, as it couples to them directly via a well known cross-section. This stands in contrast to optics, since the theoretical description of the optical cross-section for these modes is an area of active research \cite{Louie2021discovering}. 

In our model, the hopping parameters for the simulation were fixed using first-principles calculations. Since the effective octahedral crystal field parameter $10Dq$, Hund's coupling $J_H$, and the charge-transfer energy $\Delta$, depend sensitively on electron-electron interactions, we follow typical analysis approaches for \gls*{RIXS} \cite{Haverkort2012multiplet} and treated these as free parameters to fit the data. We also adjusted the crystal field levels of the individual $t_{2g}$ orbitals to reproduce the splitting seen in the lowest energy manifold (1.4-1.7~eV). The results of the model shown in Fig.~\ref{fig3} represent the best match to the intensities at this momentum transfer.

The results of this modeling are shown in Fig.~\ref{fig3}(b), along with the measured spectra in Fig.~\ref{fig3}(a). The model can reproduce the energies of the main three features [labeled 1,2, and 3 in Fig.~\ref{fig3}(a)] at $\sim$1.6, 1.8, and 2.3~eV energy loss. The position of peak 1 was found to depend primarily on $10Dq$, and peak 2 on $J_H$. Once these two parameters were fixed, the value of $\Delta$ was adjusted to give the best match for the energy of peak 3. The resultant parameters are all physically reasonable. The Coulomb interactions correspond to a 70\% screening with respect to atomic values, which is typical for transition metals \cite{degrootbook}. The small value for the charge transfer energy reaffirms \gls*{ARPES} and \gls*{DFT} predictions of strong mixing between Cr and S/Br at the Fermi level, \cite{Watson2024giant}, and the crystal field values are of a comparable magnitude to other \gls*{vdW} materials \cite{He2025dispersive, He2024magnetically, Kang2020coherent, Son2022Multiferroic, Occhialini2024Nickel, Occhialini2025spinforbidden}.

The ground state of the model is high-spin $S\approx3/2$ with strong ligand character and mixing between $d^3$, $d^4\underline{L}$, and $d^5\underline{L^2}$ configurations, where $\underline{L}$ denotes a ligand hole. The different excitations can be identified by computing the orbital occupancies, as well as the spin and orbital magnetic moment expectation values for our \gls*{AIM}. These results are shown in Fig.~\ref{fig3}(c) and (d), showing that the lowest energy manifold of excitations has a primarily $t_{2g}\rightarrow e_g$ character. This is indicated by the change in occupancy of the $t_{2g}$ and $e_g$ levels compared to the ground state, with no change in the spin or orbital moments. While we cannot unambiguously identify the individual sub-peaks within feature 1, all excitations within this energy range have $t_{2g}\rightarrow e_g$ character. Feature 2 has instead a Hund's character, where $t_{2g}$ occupancy remains the same but total spin is reduced. The highest energy manifold of excitations (peak 3) also has $t_{2g}\rightarrow e_g$ character, but with a reduced value of the total orbital angular momentum. This reflects the increased contribution of the $t_{2g}\rightarrow e_g$ transitions with higher Coulomb energy cost. 

The predicted character of the lowest energy, dark dispersive exciton is of particular interest. In view of the dispersive nature of the low energy modes, it is clear that they have particularly extended character, so they cannot be completely described by the cluster model. Our modeling nonetheless indicates that the low-energy excitons must have $t_{2g} \rightarrow e_g$ character. This is notable as it distinguishes CrSBr from some other excitons in \gls*{vdW} materials such as NiPS$_3$, which hosts an exciton that involves a high-spin to low-spin transition within the same orbital manifold \cite{Kang2020coherent, He2024magnetically}. In order to move the spin-flip excitation down to the observed energy of the dispersive exciton, the J$_H$ parameter must be reduced to unphysically small values where the Slater (Coulomb interaction) parameters are scaled down to approximately half of their atomic values. This is below the range of values that are considered physically reasonable \cite{degrootbook}. 

\textit{Conclusions.}---Our \gls*{RIXS} measurements of electronic excitations in CrSBr allowed us to characterize the momentum dependence of the full complement of electronic excitations, including optically dark excitons not previously reported. The co-existence of \gls*{2D} dark excitons and \gls*{1D} bright excitons may open long-term routes to control the directionality of exciton transport by exciting CrSBr at different energies \cite{Malic2023exciton} and dark excitons offer possibilities for modifying the optical activity of these excitons in search of future coherence applications \cite{Mueller2018, Molas_2017, Zhu2024}. Comparing the measured spectra with the model results also allows us to determine the essential orbital and spin character of these excitons, providing a comprehensive characterization of the electronic excitations in this candidate functional material.

We detect the previously reported bright exciton at 1.38(1)~eV, but also find a rich spectrum including a number of dark excitons at higher energies. The bright exciton shows the momentum dependence expected for a highly one-dimensional band-edge excitation, while most of the dark excitons do not show dispersion. However, the lowest energy dark exciton (1.46~eV at the Brillouin zone center and 1.4~eV at $\vec{q}=(-0.29, 0, 0.28)$) shows more exotic behavior namely a strong temperature dependence and an anisotropic dispersion. This excitation disperses downwards along $H$ and upwards along $K$, both with similar $\sim~50$~meV bandwidth. Such a dispersion is not compatible with a conventional Wannier exciton, but suggests that this excitation propagates via exchange processes.

Our simulations using a cluster model allow us to assign the character of the various observed excitons. We find that the lowest energy ($<1.8$~eV) excitations involve excitation of $t_{2g}$ electrons to the $e_g$ levels. The Hund's coupling has a slightly higher energy scale, leading to a peak in the \gls*{RIXS} spectrum at 1.9~eV, where the excited state involves a spin flip of one of the $t_{2g}$ electrons. We conclude that the anisotropic dispersing dark exciton we observe at 1.46~eV most likely has this $t_{2g}\rightarrow e_g$ orbital character and does not involve a spin flip. The discovery of this unusual exciton underlines the coexistence of \gls*{1D} and \gls*{2D} behavior in CrSBr.

\begin{acknowledgments}
Work at Brookhaven is supported by the Office of Basic Energy Sciences, Materials Sciences and Engineering Division, U.S.\ Department of Energy (DOE) under Contract No.\ DE-SC0012704. Work at the University of Tennessee (RIXS calculations and interpretation by model Hamiltonian calculations) was supported by the U.S.\ Department of Energy, Office of Science, Office of Basic Energy Sciences, under Award No.~DE-SC0022311. E.B. was supported by the United States Army Research Office (W911NF-23-1-0394). This research used \gls*{ESRF} beam line ID32 under the proposals HC5030. Part of this research (T.B.) was conducted at the Center for Nanophase Materials Sciences, which is a DOE Office of Science User Facility. The work by J.W.V.\ is supported by the Quantum Science Center (QSC), a National Quantum Information Science Research Center of DOE. We also acknowledge resources made available through BNL/LDRD\#19-013. This research used beamline 2-ID of the National Synchrotron Light Source II, a U.S.\  DOE Office of Science User Facility operated for the DOE Office of Science by Brookhaven National Laboratory under Contract No. DE-SC0012704. Crystal structures were rendered using VESTA \cite{vesta}.
\end{acknowledgments}

\bibliography{refs}

\appendix

\section{RIXS calculations}

The CrSBr data were interpreted using \gls*{ED} methods based on the Kramers-Heisenberg equation \cite{Mitrano2024exploring, Wang2019EDRIXS, EDRIXS}. Due to the strongly correlated nature of CrSBr, it is important to treat electron-electron interactions accurately \cite{Watson2024giant}. For these reasons, cluster approximations are particularly appropriate for modeling \gls*{RIXS}  \cite{Ament2011resonant, Mitrano2024exploring} and have been successfully applied to studying excitons in a number of other \gls*{vdW} systems \cite{He2025dispersive, He2024magnetically, Kang2020coherent, Son2022Multiferroic, Occhialini2024Nickel, Occhialini2025spinforbidden}. We therefore use an \gls*{AIM} constructed from the cluster shown in Fig.~\ref{fig1}(a) in which the Cr atoms are surrounded by 4 S and 2 Br atoms. 

\begin{figure}[htb!]
    \centering
    \includegraphics[width=0.4\textwidth]{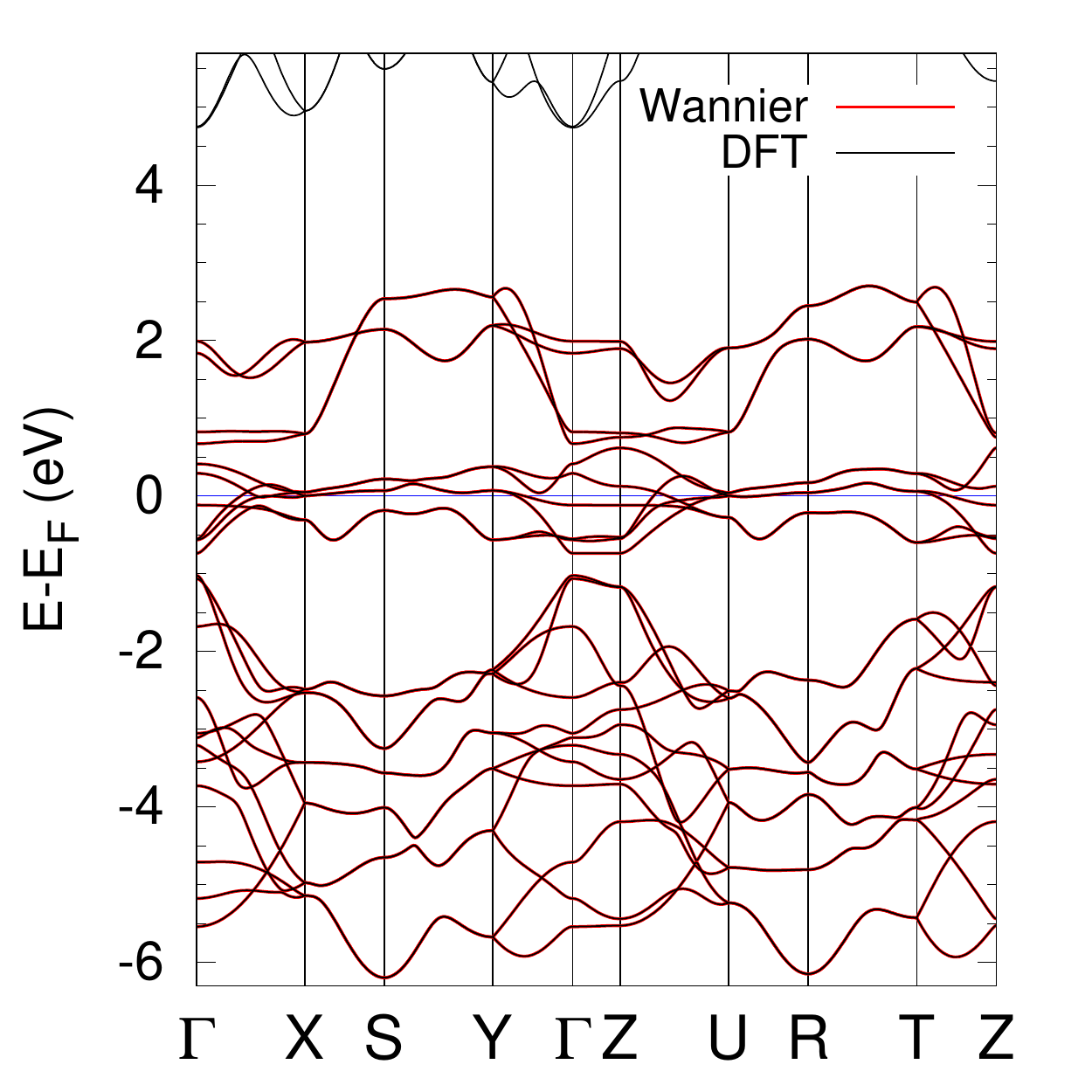}
    \caption{Band structure of CrSBr without spin-orbit coupling. DFT calculated bands are in black, overlain on the Wannier projected bands in red, showing their total agreement. The symmetry labels follow the conventional labeling methodology described in Ref.~\cite{Setyawan2010high} based on the Bravais lattice of the unit cell.}
    \label{fig:bandstructure}
\end{figure}

To determine the hopping parameters for our model, we first computed the electronic structure of CrSBr using the \gls*{DFT} code \textsc{VASP} \cite{Kresse1996effcient, Kresse1996effciency}. The calculations are performed within the Perdew-Burke-Ernzerhof generalized gradient approximation \cite{GGA} for the exchange-correlation functional without spin-orbit coupling. We use projector augmented wave pseudopotentials \cite{PAW1,PAW2} with an energy cutoff of 450 eV and a $20 \times 14 \times 10$ Monkhorst-Pack $k$-point mesh.  We constructed a tight-binding model using \textsc{Wannier90} \cite{Mostofi2014updated, Marzari1997maximally, Souza2001maximally} by performing a Wannier projection of Cr $3d$, S $3p$, and Br $4p$ orbitals without maximal localization; as these constitute an isolated manifold of bands, disentanglement is also unnecessary. Figure~\ref{fig:bandstructure} shows an excellent agreement between the DFT and Wannier function band structures. Since only a subset of the $p$ orbitals bond with Cr $d$ states, the model can be implemented much more efficiently with essentially no loss in accuracy by representing the ligand states by 5 effective ligand orbitals that hybridize most strongly with the central Cr site \cite{Haverkort2012multiplet}. 


The $x$, $y$ and $z$ coordinate system used to describe the atomic and molecular orbitals, and perform the \gls*{RIXS} calculation, do not coincide with the crystallographic $a$, $b$ and $c$ axes. This was done to obtain coordinates that are approximately aligned with the bond directions of the distorted octahedron in CrSBr and simplify the interpretation of the molecular orbitals. The $z$ axis is along the crystallographic $b$ axis, while $x$ and $y$ are in the $a-c$ plane and $45^{\circ}$ rotated compared to the crystallographic axes. $x$ is in the $a,c>0$ quadrant, and $y$ is in the $a>0,c<0$ quadrant.

The Hamiltonian describing the cluster includes a four-fermion Coulomb term, and a two-fermion term which includes contributions from crystal field, hopping, spin-orbit, and magnetic field. The diagonalization of the Hamiltonian, and calculation of the cross sections were performed using the Fortran \gls*{AIM} solver provided by \textsc{EDRIXS} package \cite{Wang2019EDRIXS, EDRIXS} following an approach similar to that used for NiPS$_3$ and CrI$_3$ \cite{He2024magnetically, He2025dispersive}.\\

 The four-fermion Coulomb term of the Hamiltonian includes interactions within the Cr $3d$ orbitals, as well as between the Cr $3d$ and Cr $2p$ orbitals in the intermediate state where a core hole is present. Coulomb interactions involving the ligand orbitals are expected to be relatively weak and so were not included. The Coulomb interactions for the Cr $3d$ orbitals can be described in terms of the Slater parameters $F^0_{dd}$, $F^2_{dd}$, and $F^4_{dd}$, while those between the Cr $3d$ and Cr $2p$ are parametrized by $F^0_{dp}$, $F^2_{dp}$, $G^1_{dp}$, $G^3_{dp}$. Of these, $F^0_{dd}$ and $F^0_{dp}$ depend on the on-site Cr $3d$ Coulomb parameters $U_{dd}$ and $U_{dp}$, which were set to typical values of 4 eV and 6 eV respectively \cite{Bocquet1992systematics, Feldkemper1998generalized}. The results of the simulation do not depend strongly on the precise values chosen. The rest of the parameters ($F^2_{dd}$, $F^4_{dd}$, $F^2_{dp}$, $G^1_{dp}$, and $G^3_{dp}$) were set to their Hartree-Fock atomic values calculated by Cowan's code \cite{Cowan1981theory}, scaled down to account for screening effects in the crystal. One scaling factor was used for the Cr $3d$ orbitals and one for the core-hole interactions, denoted $k_{dd}$ and $k_{dp}$ respectively. $k_{dp}$ does not strongly affect the fit to the experimental data, and so is fixed to $0.7$. This leaves the scaling parameter $k_{dd}$, which is proportional to the Hund's coupling, as the only free parameter in the Coulomb portion of the Hamiltonian. All of the Coulomb parameters are listed in Table~\ref{table:four_fermion_terms}. \\

\begin{table*}
\caption{Input parameters for the four-fermion term. All numbers are in units of eV, with the exception of $k_{dd}$ and $k_{dp}$ which are scaling parameters.}
\begin{ruledtabular}
\begin{tabular}{cccccccccccc}
\hline\\
\multicolumn{12}{c}{Initial State}\\
\multicolumn{5}{c}{$d$-$d$ Coulomb interactions} & & \multicolumn{6}{c}{ }\\
\hline\\
$U_{dd}$   & $k_{dd}$ & $F^0_{dd}$   & $F^2_{dd}$   & $F^4_{dd}$  &&&&&&&\\
4 & \textbf{0.72} & 4.4 & 7.76 & 4.86 &&&&&&&\\

\hline\\
\multicolumn{12}{c}{Intermediate State}\\
\multicolumn{5}{c}{$d$-$d$ Coulomb interactions} & & \multicolumn{6}{c}{$d$-$p$ Coulomb Interactions}\\
$U_{dd}$   & $k_{dd}$ & $F^0_{dd}$   & $F^2_{dd}$   & $F^4_{dd}$  & &$U_{dp}$   & $k_{dp}$ & $F^0_{dp}$   & $F^2_{dp}$   & $G^1_{dp}$ & $G^3_{dp}$\\
4 & \textbf{0.72} & 4.43 & 8.35 & 5.23 &     &  6 & 0.7 & 6.31 & 4.57 & 3.35 & 1.91\\
\end{tabular}
\end{ruledtabular}
\label{table:four_fermion_terms}
\end{table*}


The two-fermion term includes contributions from crystal field, hopping, spin-orbit coupling, and magnetic field. The crystal field and hopping terms are derived from the \gls*{DFT} calculation, and are shown in Eq.~\ref{eq:hopping} as a $10\times10$ matrix. The order of the orbitals is as shown, listing first the Cr $d$ orbitals and then the ligand $L$ orbitals. The electronic structure calculation was not spin-resolved, and the same values were used for both spin-up and spin-down states in our \textsc{EDRIXS} calculation.  \\

\begin{widetext}
\begin{equation*}\label{eq:hopping}
\begin{pNiceMatrix}[first-row,first-col] 
         &d_{x^2-y^2}&d_{3z^2-r^2}&d_{xy}&d_{xz}&d_{yz}&L_{x^2-y^2}&L_{3z^2-r^2}&L_{xy}&L_{xz}&L_{yz} \\
d_{x^2-y^2}& 3.201&0&0&0&0&-2.426&0&0&0&0 \\
d_{3z^2-r^2} &0&3.227&0.0147&0&0&0&-2.313&-0.15&0&0 \\
d_{xy} &0&0.0147&2.662&0&0&0&0.159&-1.49&0&0 \\
d_{xz} &0&0&0&2.733&-0.041&0&0&0&-1.383&-0.068 \\
d_{yz} &0&0&0&-0.041&2.733&0&0&0&-0.068&-1.383 \\
L_{x^2-y^2} &-2.426&0&0&0&0&1.573&0&0&0&0 \\ 
L_{3z^2-r^2} &0&-2.313&0.159&0&0&0&1.56&0.211&0&0 \\ 
L_{xy} &0&-0.15&-1.49&0&0&0&0.211&-0.181&0&0 \\
L_{xz} &0&0&0&-1.383&-0.068&0&0&0&0.0096&-0.061 \\
L_{yz} &0&0&0&-0.068&-1.383&0&0&0&-0.061&0.0096
\end{pNiceMatrix}
\end{equation*}
\end{widetext}

Since hopping depends on how electronic wavefunctions are spread between different atoms, it is only weakly influenced by strongly correlated physics and \gls*{DFT} generally captures the magnitude of hopping reasonably accurately. For this reason, we consider the off-diagonal inter-orbital hopping values to be good estimates for the true values. However it was necessary to adjust the diagonal (crystal field) terms as \gls*{DFT} includes Coulomb interactions at the mean-field level, so the on-site energies are subject to double-counting errors, which we need to avoid. In order to fit the data we therefore add a further splitting between the Cr $e_g$ and $t_{2g}$ orbitals, denoted $\Delta10D_q$. The energy levels of the Cr and ligand states are also adjusted by adding an offset $E_d$ to the diagonal terms for the metal orbitals and $E_L$ to those for the ligand orbitals. These offsets were calculated using the \textsc{EDRIXS} functions \textbf{edrixs.utils.CT\_imp\_bath} and \textbf{edrixs.utils.CT\_imp\_bath\_core\_hole} with the arguments $U_{dd}=4$ and $\delta=0$. Adjusting this parameter $\delta$ has the effect of tuning the charge transfer energy $\Delta_\text{ct}$, which is defined as the cost to move a hole from the metal $d$ orbitals onto the lowest energy ligand orbital. We note that $\delta$ and $\Delta_\text{ct}$ are not related in a straightforward way. $\Delta_\text{ct}$ was calculated by diagonalizing the Hamiltonian with Cr-ligand hopping set to zero and then taking the energy difference between the $d^3$ and $d^4\underline{L}$ states (where $\underline{L}$ denotes a ligand hole).\\ 

\begin{table*}
\caption{Additional input parameters for two-fermion term. All numbers are in units of eV.}
\begin{ruledtabular}
\begin{tabular}{cccccccccc}
Charge transfer  & \multicolumn{3}{c}{Crystal field}  &  \multicolumn{3}{c}{Spin-orbit coupling} & Magnetic field  \\
\hline\\

$\Delta_\text{ct}$   &  $\Delta 10D_q$ & $\Delta d_{xy}$ & $\Delta d_{xz}/d_{yz}$ mixing & $\zeta_i$  & $\zeta_n$   & $\zeta_c$ & $H$   \\
\textbf{0.8} &  \textbf{0.23} & \textbf{0.27} & \textbf{-0.06} &0.035  & 0.047  & 5.667  &  0.015  \\
\end{tabular}
\end{ruledtabular}
\label{table:two_fermion_terms}
\end{table*}

The crystal field levels of the $t_{2g}$ orbitals were also adjusted to reproduce the splitting in the lowest energy part of the experimental spectra. Since all of the $d$ levels are in principle independent due to the low symmetry, there are multiple parameter choices which can place peaks at the energies matching the experimental spectra. The parameter choice which we found best matched the peak intensities was obtained by adjusting only the $t_{2g}$ levels. The $d_{xy}$ energy was shifted up by 0.27~eV and the off-diagonal mixing term for $d_{xz}$ and $d_{yz}$ was decreased by 0.06~eV. We note, however, that other parameter sets may also give relatively good fits to the available experimental data. Regardless of the detailed parameterization, the lowest energy exciton around 1.38~eV appears due to the splitting of a single orbital down to lower energy, which occurs due to the very low symmetry of CrSBr.\\

Spin-orbit coupling was also included in the two-fermion term, though this interaction is weak and does not strongly affect the results of the simulation. The spin-orbit parameters were therefore taken to be the atomic values. A small magnetic field of 15~meV, which splits the high spin $S\approx3/2$ ground state into four non-degenerate levels, was introduced to mimic the environment in the ferromagnetically ordered layers (the magnitude of the field was selected based on the magnon bandwidth reported in neutron measurements)\cite{Scheie2022spin}. As a result, there were two free parameters in the two-fermion term: the $10D_q$ splitting of the Cr $d$ orbitals, and the splitting between the metal and the ligand orbitals parametrized as $\Delta_\text{ct}$. \\

In summary, our calculation used three free parameters to fit the main features: an offset to the octahedral crystal field splitting $\Delta10Dq$, the Coulomb scaling parameter $k_{dd}$ (which scales with the Hund's coupling), and the charge transfer energy ($\Delta_\text{ct}$). These parameters (indicated in bold in Tables \ref{table:four_fermion_terms} and \ref{table:two_fermion_terms}) were adjusted to best match the experimental \gls*{RIXS} spectra, while all other parameters were fixed to the specified values. Two additional parameters were then used to adjust the splitting of the $t_{2g}$ orbitals to better match the finer structure in the low energy peak, however we were not able to fully constrain these parameters based on the available data. Once the Hamiltonian was diagonalized, the \gls*{RIXS} and \gls*{XAS} cross sections were calculated for the experimental geometry and polarization settings used for the measurements. An inverse core-hole lifetime broadening of $\Gamma_c = 0.6$ eV was used to compute the \gls*{RIXS} spectra. We used a final state inverse lifetime of $\Gamma_f = 0.03$ eV, which has the effect of broadening the spectra by a Lorentzian function with a full width half max of 0.03 eV.

\end{document}